\documentclass[12 pt]{article}
\usepackage[utf8]{inputenc}
\usepackage{tikz,tkz-euclide,pgfplots}
\usepackage{comment}
\usepackage{textcomp}
\usepackage{bbm}
\usepackage{cancel}
\usepackage{physics}
\usepackage{appendix}
\usepackage{biblatex} 
\usepackage{amsthm}
\usepackage{amsmath}
\usepackage{amsfonts}
    \usetikzlibrary{arrows,calc,patterns}
    \usepackage{mathtools}
    \usepackage{authblk}
\usepackage{hyperref}
\hypersetup{
    colorlinks=true,
   linkcolor=blue,
    filecolor=magenta,      
   urlcolor=cyan,
}

\title{Where is the photon in an interferometer?}
\author{Joseph E. Avron}
\affil {Faculty of Physics,\\ Technion, Haifa, Israel}
\date{\today}
\usepackage{graphicx}

\pgfplotsset{compat=1.14}
\begin{document}
\maketitle
\begin{abstract}
In a paper titled ``Asking photons where they have been?" \cite{LV}  Danan, Farfurnik, Bar-Ad and Vaidman describe an experiment with pre and post selected photons going through nested  Mach-Zehnder interferometers. They find that some of the mirrors leave no footprints on the signal and interpret this as evidence that the  photon skipped these mirrors. They argue that the experiment supports Aharonov-Vaidman's formulation of quantum mechanics \cite{AharonovVaidman} where post-selected particles are assigned disconnected trajectories. I review the experiment and analyze it within the orthodox framework of quantum mechanics. The standard view of interfering  trajectories accounts for the experimental findings.\end{abstract}
\hfill
\begin{minipage}[r]{.5\textwidth}
\it{Alex Grossmann has been a beacon of light and warmth to me, a teacher, mentor, dear friend and father figure.}
\end{minipage}
\section{The experiment}

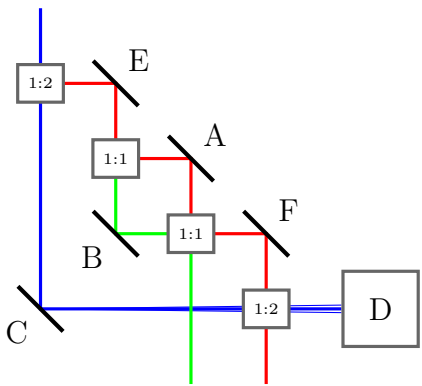
\begin{figure}
    \centering
    \begin{tikzpicture}[scale=3/3,squarednode/.style={rectangle, draw=black!60, very thick,fill=white, minimum size=5mm},dsquarednode/.style={rectangle, draw=black!60, very thick, minimum size=10mm}]
    
\draw[red, very thick] (-1,1) -- (0,1) -- (0,0) -- (1,0) -- (1,-1) -- (2,-1) --(2,-3);
\draw [blue, very thick] (-1,2) -- (-1,-2)-- (3,-2);
\draw [blue, thin]  (-1,-2)-- (3,-2.05);
\draw [blue, thin]  (-1,-2)-- (3,-1.95);
\draw [green, very thick] (0,0)-- (0,-1) -- (1,-1) -- (1,-3);
\draw [ultra thick] (1-.3,0+.3) --(1+.3,0-.3) node [above right] at (1,0) {A};
\draw [ultra thick] (0-.3,1+.3) --(0+.3,1-.3) node [above right] at (0,1) {E};
\draw [ultra thick] (2-.3,-1+.3) --(2+.3,-1-.3) node [above right] at (2,-1) {F};
\draw [ultra thick] (-1-.3,-2+.3) --(-1+.3,-2-.3) node [below left] at (-1,-2) {C};
\draw [ultra thick] (0-.3,-1+.3) --(0+.3,-1-.3) node [below left] at (0,-1) {B};
\node[squarednode]  at (0,0) {\tiny{1:1}};
\node[squarednode]  at (-1,1) {\tiny{1:2}};
\node[squarednode]  at (1,-1) {\tiny{1:1}};
\node[squarednode]  at (2,-2) {\tiny{1:2}}; 
\node[dsquarednode,right] at (3,-2) {D};
\end{tikzpicture}
    \caption{A nested Mach-Zehnder interferometer: The four squares represent beam-splitters where the ratio $p\!:\!q$ give the intensity ratio of the outgoing beams. The external Mach-Zehnder has $1\! :\! 2$ beam splitters and the internal $1:1$ beam splitters. The black lines marked \{A,\dots, E\} represent mirrors that slightly oscillate, each with its own specific frequency. Light is fed in the upper blue port and  measured by a ``quad-detector" $D$ which compares the signal on its top half with the bottom half. The colors correspond to the colors of the modes in Fig.~\ref{fig:YA}}
    \label{fig:LV}
\end{figure}

Consider the  nested Mach-Zehnder interferometers shown in Fig.~\ref{fig:LV}.
Each one of the five  mirrors $\{A,B,C,E,F\}$ oscillates with its characteristic frequency. The intensity of light falling on the top half of the detector surface is compared with the intensity falling on the bottom half. The power spectrum of the signal bears evidence to the oscillation frequencies of the  mirrors.

The interferometer is tuned so that  frequencies corresponding to mirrors $A,B,C$ show up in the power spectrum but those of $E,F$ do not.
The appearance of a characteristic frequency in the signal gives, of course, evidence that photons hit the corresponding mirror. The authors of \cite{LV} go one step further and interpret the absence of a characteristic frequency as evidence that no photon visited the corresponding mirror.  Since the photon apparently succeeded in going undetected by the gate keepers of the $A,B$ mirrors, it seemed to have a skipping orbit that visited disconnected parts of space.
\section{The history of post-selected states}

\begin{figure}
    \centering
    \begin{tikzpicture}[scale=3/3,squarednode/.style={rectangle, draw=black!60, very thick,fill=white, minimum size=5mm},dsquarednode/.style={rectangle, draw=black!60, very thick, minimum size=10mm}]
\begin{scope}[shift={(0.07,.07)}]
 \draw[red,  thin,<-] (-2,0) -- (1,0) -- (1,-1) -- (2,-1) --(2,-2);
\draw [red,  thin] (-1,1.93) -- (-1,-2)-- (3-.07,-2);
\draw [red,  thin] (0,0)-- (0,-1) -- (1,-1);
 \end{scope}
 
 \begin{scope}
\draw[blue, very thick] (-1,1) -- (0,1) -- (0,0) -- (1,0) -- (1,-1);
\draw [blue, very thick] (-1,2) -- (-1,-2)-- (3,-2);
\draw [blue, very thick,->] (0,0)-- (0,-1) -- (1,-1) -- (1,-3);
\draw [ultra thick,shift={(.07,.07)}] (1-.3,0+.3) --(1+.3,0-.3) node [above right] at (1,0) {A};
\draw [ultra thick] (0-.3,1+.3) --(0+.3,1-.3) node [above right] at (0,1) {E};
\draw [ultra thick, shift={(.07,.07)}] (2-.3,-1+.3) --(2+.3,-1-.3) node [above right] at (2,-1) {F};
\draw [ultra thick] (-1-.3,-2+.3) --(-1+.3,-2-.3) node [below left] at (-1,-2) {C};
\draw [ultra thick] (0-.3,-1+.3) --(0+.3,-1-.3) node [below left] at (0,-1) {B};
\node[squarednode]  at (0,0) {\tiny{1:1}};
\node[squarednode]  at (-1,1) {\tiny{1:2}};
\node[squarednode]  at (1,-1) {\tiny{1:1}};
\node[squarednode]  at (2,-2) {\tiny{1:2}}; 
\node[dsquarednode,right] at (3,-2) {D};
 \end{scope}
 
\end{tikzpicture}
    \caption{The blue graph describe the state that evolve forward in time and the red graph the state that starts at the detector and evolve backward in time. The blue graph is a connected set and so is the red graph. However, their intersection is a disconnected set: The line through C  and the square touching A and B.    }
    \label{fig:loops}
\end{figure}
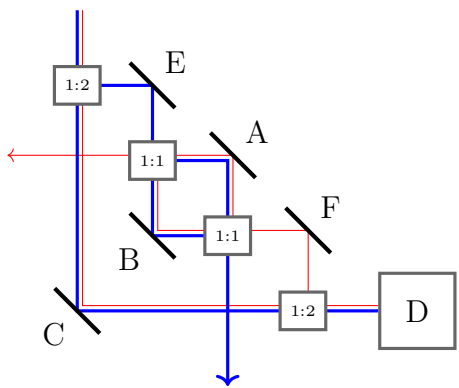

The two state formulation of Aharonov and Vaidman \cite{AharonovVaidman} attempts to assign a consistent history to  post selected states. The input state in Fig.~\ref{fig:loops} evolves forward in time growing to the blue graph.  The post-selected state starts at the detector and evolves backward in time to become the red graph.  The consistent history is, by definition, their intersection which  is the disconnected set, the $L$ shaped line and the square  in Fig.~\ref{fig:loops}.  

The notion of consistent history gives a simple account of the experiment of Danan et.~al.~\cite{LV}: The photon never visited the E and F mirrors.  However, the explanation comes with the price tag of disconnected trajectories. This is not what textbooks in quantum mechanics teach students. 

Danan et.~al.~\cite{LV} are careful not claim that a disconnected path is the inevitable conclusion of their experiment. In fact, they show that the experiment can also be explained in terms of interfering classical waves. However, there is little doubt that the attention drawn to the work owes much to the support it gives to  the  unorthodox approach of Aharonov and Vaidman \cite{AharonovVaidman} and the concomitant controversy \cite{Controversy} .  

The  experiment has been analysed in \cite{LV} both within Aharonov and Vaidman unorthodox formulation of quantum mechanics  and also within classical Maxwell theory. What still appears to be missing is its analysis within  orthodox quantum mechanics \cite{AP}. This is what I  do here. The analysis is based on a quantum circuit model. I recover the main results in \cite{LV}, and explain the weak footprints of some of the mirrors in terms of destructive quantum amplitudes.


\section{A  quantum circuit model}
The orientations of the mirrors $A,B,C,E,F$ are functions of time. As the interaction time of the photon with each mirror is very short,  one may consider the mirrors to be effectively at rest. The angles of the mirrors  may then be viewed as parameters.  The measurement of the power spectrum in \cite{LV} is, for the purpose of the analysis, just a clever trick to gain information about the instantaneous angles $\alpha,\beta,\gamma,\eta,\phi$, of the mirrors.   
\begin{figure}
    \centering
    \begin{tikzpicture}[scale=.9,squarednode/.style={rectangle, draw=black!60,fill=white, very thick, minimum size=1.8cm},ssquarednode/.style={rectangle, draw=black!60,fill=white, very thick, minimum size=.6 cm}]
  \node [left] at (-5,1) {$\ket\varphi$}; 
      \node [left] at (-5,0) {$\ket 0$}; 
            \node [left] at (-5,-1) {$\ket 0$}; 
\node [above] at (-4,1) {$1$};
\node [above] at (-4,0) {$2$};
\node [above] at (-4,-1) {$3$};
\node [above] at (7,1) {$1$};
\node [above] at (7,0) {$2$};
\node [above] at (7,-1) {$3$};

\draw[blue, very thick] (-5,1) --(-4,1)--(-3,1/2)--(-2,1)--(-1,1) --(-4+9,1)--(-3+9,1/2)--(-2+9,1)-- (8.5,1);
\draw [red, very thick] (-5,0) --(-4,0)--(-3,1/2)--(-2,0)-- (-1,0) --(0,-1/2) --(1,0)-- (-1+3,-1+1) --(0+3,-1/2) --(1+3,-1+1)--(5,0)--(6,1/2)--(7,0)--(8.5,0);
\draw [green, very thick] (-5,-1) --(-1,-1) --(0,-1/2) --(1,-1)-- (-1+3,-1) --(0+3,-1/2) --(1+3,-1)--(8.5,-1); 
\draw [opacity=.5, blue] (1.5,1)--(20:5) node [above right] at (3.5,.95) {$\gamma$};
\draw [opacity=.5, blue] (4.5,0)--(6.5,-.5) node [above right] at (5.5,-1/2) {$\phi$};
\node[ssquarednode] at (-1.5,0) {E};
\node[ssquarednode] at (1.5,1) {C};
\node[ssquarednode] at (4.5,0) {F};
\node[ssquarednode] at (1.5,0) {A};
\node[ssquarednode] at (1.5,-1) {B};
\node[ssquarednode,right] at (7.8,1) {$D$};
\draw [dashed] (-.5,-1.5) rectangle (3.5,0.5);
\node  at (-3,1) {$S_{12}$};
\node  at (6,1) {$S_{12}$};
\node  at (0,-1.) {$S_{23}$};
\node  at (3,-1.) {$S_{23}$};
\end{tikzpicture}

    \caption{A circuit diagram corresponding to the experimental setting in Fig.~\ref{fig:LV}. $A,B,C,E,F$ represents the mirrors with deflection angles $\alpha,\beta,\gamma,\eta,\phi$.  The deflection angles $\gamma$ and $\phi$ are shown. The beam splitters are represented by the crossings. $D$ is the detector. The three channels are markes $1$ (blue), $2$ (red) and $3$ (green). $\ket 0$ denotes the photonic vacuum and $\ket{\varphi}$ a single photon state. }
    \label{fig:YA}
\end{figure}
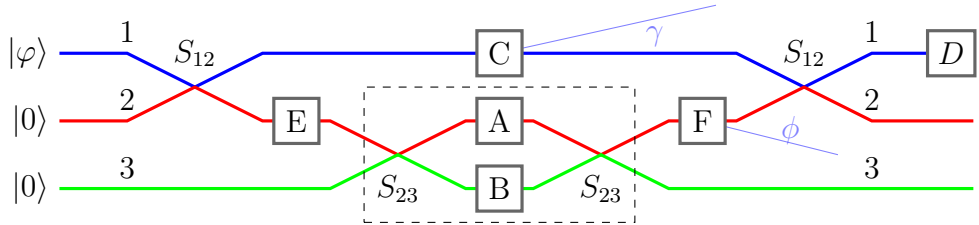

A quantum mechanical model corresponding to experiment is the circuit of gates shown in Fig.~\ref{fig:YA}. The photon (annihilation) operator has two coordinates: The channel index $j$ and wave-vector $\mathbf{k}$:
\begin{equation}
    a_{j}(\mathbf{k}), \quad j\in \{1,2,3\},\quad \mathbf{k}\in \mathbbm{R}^2 
\end{equation}
(The three channels are marked blue, red and green in Fig.~\ref{fig:YA} and $\mathbf{k}$ describes the photon wave vector in the plane.)  
The modes satisfy the canonical commutation relations
\begin{equation}
    [a_{j}(\mathbf{k})^{\phantom{\dagger}},a_{k}(\mathbf{k}')^\dagger]=\delta_{jk}\delta(\mathbf{k}-\mathbf{k}')
\end{equation}

The beam splitter $S_{jk}$ acts trivially on   $\mathbf{k}$ and non-trivially on the mode indices $j,k$. Since a beam splitter is time-reversal invariant the corresponding matrix may be chosen symmetric\footnote{A scattering matrix $S$ is time reversal invariant if $T S=S^{-1}T$ with $T$ the anti-unitary time reversal. Gauge freedom allows to redefine $S_{jk}\mapsto z_j S_{jk} \bar w_k$ where $|z_j|=|w_k|=1$ are arbitrary phases assigned to the incoming and outgoing modes. 
}. The two beam splitters may be chosen as:
\begin{equation}\label{e:s}
    S_{12}(p:q)=\begin{pmatrix}\sqrt p &i\sqrt q\\
i\sqrt q  &\sqrt p
\end{pmatrix}, \quad S_{23}=\frac 1  { \sqrt 2}\begin{pmatrix} 1 &i\\
i  & 1
\end{pmatrix} 
\end{equation}
with $p+q=1$. $p:q$ is the intensity ratio of the outgoing beams. In Fig.~\ref{fig:LV} $p\mapsto 1, q\mapsto 2$. 

The mirrors act trivially on the channel index $j$ and non-trivially on the wave vector  $\mathbf{k}$.
The mirror $A$, with angle $\alpha/2$ (deflection angle $\alpha$), acts on the vector  $\mathbf{k}$  as a reflection, represented by the orthogonal, symmetric  matrix  $M_\alpha$ with $\det M_\alpha=-1$:
\begin{equation}\label{e:mirror}
    \mathbf{k}\mapsto M_\alpha \mathbf{k}, \quad M_\alpha=\begin{pmatrix}
    \cos\alpha & \sin\alpha\\
    \sin\alpha& -\cos\alpha
    \end{pmatrix}
    ,\quad M_\alpha^2=\mathbbm{1}
\end{equation}
The mirror acts on $\ket{\mathbf{k}}$ as a unitary map,
\begin{equation}\label{e:u-mirror}
U_A=\int d \mathbf{k} \ket{M_\alpha\mathbf{k}}\bra{\mathbf{k}}
\end{equation}
Similarly for the other mirrors.

Since a reflection is an orthogonal transformation that is orientation reversing, an even number of reflections is a rotation and an odd number is a reflection. The total deflection associated with the path $E-A-F$ is the reflection
\begin{equation}\label{e:ta}
M_\phi M_\alpha M_\eta= M_{\tilde\alpha}, \quad \tilde\alpha= -\alpha+\eta+\phi
\end{equation}
Similarly for the path $E-B-F$
\begin{equation}\label{e:tb}
M_\phi M_\beta M_\eta= M_{\tilde\beta}, \quad \tilde\beta= -\beta+\eta+\phi
\end{equation}
One can decorate the circuit with unitaries that represent  phase delays between channels. This is important in practice as it allows to tune the interferometers.  However, for  the sake of simplicity it is best not to.

\section{Three interfering paths}

A photon entering the circuit at the blue port $1$ of Fig.~\ref{fig:YA} has three routes  to get to the detector: The  route through
$E-A-F$ and the route through $E-B-F$ and the route through $C$.

By the linearity of quantum mechanics the amplitude at the detector is the sum of the amplitudes associated with the three routes:
\begin{equation}\label{e:unitary}
   \bra{1,\mathbf{k}'}\mathbf{U}_A\ket{1,\mathbf{k}}+ \bra{1,\mathbf{k}'}\mathbf{U}_B\ket{1,\mathbf{k}}+   \bra{1,\mathbf{k}'}\mathbf{U}_C\ket{1,\mathbf{k}}
\end{equation}

The three amplitudes can be computed explicitly by tracing the gates in the three channels in Fig. \ref{fig:YA}. Since the beam splitters and mirrors act on different parts of the tensor product their action commute. And, since all the rotation angles are in the plane, they too commute.
Using Eqs.~\ref{e:s} and \ref{e:u-mirror} one finds
\begin{align}
     \bra{1,\mathbf{k}'}\mathbf{U}_A\ket{1,\mathbf{k}}&= \bra{1} S_{12}\ket{2}\bra{2}S_{23}\ket{2}^2\bra{2} S_{12}\ket{1}\delta(\mathbf{k}'-M_{\tilde\alpha}\mathbf{k})\nonumber \\ &
     = -\frac q 2\,\delta(\mathbf{k}'-M_{\tilde\alpha}\mathbf{k})\label{e:minus}
\\
     \bra{1,\mathbf{k}'}\mathbf{U}_B\ket{1,\mathbf{k}}&= \bra{1} S_{12}\ket{2}\bra{2}S_{23}\ket{3}\bra{3}S_{23}\ket{2}\bra{2} S_{12}\ket{1}\delta(\mathbf{k}'-M_{\tilde\beta}\mathbf{k})\nonumber \\ &= \frac q 2\,\delta(\mathbf{k}'-M_{\tilde\beta}\mathbf{k})\label{e:3path}\\
      \bra{1,\mathbf{k}'}\mathbf{U}_C\ket{1,\mathbf{k}}&= \bra{1} S_{12}\ket{1} \delta(\mathbf{k}'-M_\gamma\mathbf{k})=p\,\delta(\mathbf{k}'-M_\gamma\mathbf{k}) \label{e:1path}
\end{align}
The only subtlety here is the relative sign between Eq.~\ref{e:minus} and Eq.~\ref{e:3path}  which comes from $i^2=-1$ and $i^4=1$.

The first term in Eq.~\ref{e:unitary} depends on $\tilde \alpha$, the angle accumulated along the path $A$. Similarly, the second term depends on $\tilde\beta$ and the third on $\gamma$.
The detection amplitude is therefore a function of the three variables $(\tilde\alpha,\tilde\beta,\gamma)$. Since $\tilde\alpha$ and $\tilde\beta$ depend on $\eta+\phi$, variations of the angles that satisfy $\delta\eta+\delta\phi=0$ do not affect the detection. More generally, variations  of the five mirrors that satisfy \begin{equation}
    \delta \alpha =\delta(\eta+\phi), \quad  \delta \beta =\delta(\eta+\phi)
\end{equation}
do not affect the detection amplitudes at all.

\subsection{The amplitudes of the outgoing state}\label{s:amp}
Let $\ket\varphi$ be the incoming one photon state in channel $1$.
Denote the outgoing state 
\begin{equation}
    \ket{\tilde\varphi}=(\mathbf{U}_A+\mathbf{U}_B+\mathbf{U}_C)\ket\varphi
\end{equation}
Taking into account that a reflection is its own inverse, Eqs.~\ref{e:minus}-\ref{e:1path} give for the outgoing amplitude
\begin{equation}
    \braket{1,\mathbf{k}}{\tilde\varphi}= -\frac q 2 \varphi(M_{\tilde\alpha} \mathbf{k})+
     \frac q 2 \varphi(M_{\tilde\beta} \mathbf{k})+p  \varphi(M_{\gamma} \mathbf{k})
\end{equation}
When $\alpha=\beta$, then also   $\tilde\alpha=\tilde\beta$, and the first two terms cancel. The detection amplitude then depends only on the $\gamma$. A photon that goes through $E$ does not get to the detector and is dumped in the 3 (green) output channel. 

\subsection{The small parameter: 
\texorpdfstring{$\varepsilon=\alpha-\beta$}{a=b-c}}
In the case that  the mirrors $A$ and $B$ are synchronized so that $\alpha\approx\beta$ the first two terms almost cancel and to leading order in $\varepsilon=\alpha-\beta$ we have 
\begin{equation}\label{e:epsilon}
    \braket{1,\mathbf{k}}{\tilde\varphi}\approx p\,  \varphi(M_{\gamma} \mathbf{k})-\varepsilon\, \frac q 2 {\varphi'(M_{\tilde\alpha} \mathbf{k})}  \,
\end{equation}
where $\varphi'$ is the derivative of $\varphi$ with respect to the angle. This amplitude depends on $\eta+\phi$ as they appear in $\tilde\alpha$. However, 
\begin{equation}
\partial_\eta\braket{1,\mathbf{k}}{\tilde\varphi}\approx -\varepsilon\,\frac q 2\, \varphi''(M_{\tilde\alpha} \mathbf{k})
\end{equation}
is small  because of the $\varepsilon$ factor.  We conclude that when
$\alpha\approx \beta$ the detector is insensitive to the changes in the angles of the mirrors $E$ and $F$.

If we further assume that  
\begin{equation}\label{e:synch}
    \alpha\approx\beta, \quad \gamma\approx\tilde\alpha
\end{equation}
then to leading order in $\varepsilon$
\begin{equation}
    \braket{1,\mathbf{k}}{\tilde\varphi}\approx p\,  \varphi(M_{\gamma}\mathbf{k})-\varepsilon\,\frac q 2 \varphi'(M_{\gamma} \mathbf{k}) 
\end{equation}
The rhs is now independent of $\eta$ and $\phi$.  In particular, Eq.~\ref{e:synch} is the case if all the angles are small. This is the case considered in \cite{LV}.
\section{The quad-detector}

\begin{figure}
    \centering
    \begin{tikzpicture}[ssquarednode/.style={rectangle, draw=black!60,fill=white, very thick, minimum size=1 cm}]

\draw[blue, thick,->] (-5,0) --(4.4,1/2);
\draw[blue, thick,->] (-5,0) --(4.4,-1/2);
\node[ssquarednode] at (5,1/2) {$W_+$};
\node[ssquarednode] at (5,-1/2) {$W_-$};
\end{tikzpicture}

    \caption{A quad detector is a photo-detector whose output is the difference of the signal of the upper box $W_+$ from the signal of the lower box $W_-$. }
    \label{f:quad}
\end{figure}
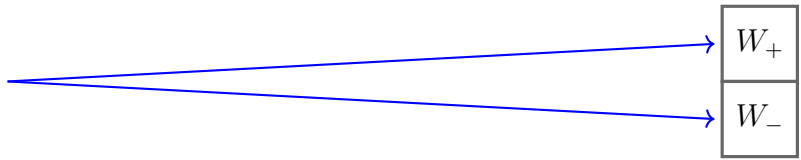

A quad-detector is a photo-detector that realizes the quantum observable of the difference of projections   
\begin{equation}
 \mathbf{I}=   \mathbf{W}_+-\mathbf{W}_-,\quad \mathbf{W}_\pm=\ket{w_\pm}\bra{w_\pm} 
\end{equation}
$w_{\pm}$ are the characteristic functions of the boxes $W_\pm$, illustrated in Fig.~\ref{f:quad}
\begin{equation}
    W_\pm=\{x,y\,|\,0>x>-1, 0<\pm y<1\}
\end{equation}

The probability amplitude for being detected by $\mathbf{W}_\pm$ is
$    \braket{w_\pm}{\tilde\varphi}$.


\subsection{Calibrating the detector}
Eq. \ref{e:epsilon} gives the detection amplitudes  
\begin{equation}
    \braket{w_\pm}{\tilde\varphi}= A_\pm+ \varepsilon\, B_\pm
\end{equation}
The interesting term $ \varepsilon B_\pm$ is masked by the dominant $A_\pm$.

The quad-detector comes with a neat calibration trick that allows to expose the subdominant term  magnifying it  using the dominant term $A_\pm$.
The signal of the quad detector is
\begin{equation}
 |\braket{w_+}{\tilde\varphi}|^2-
     |\braket{w_-}{\tilde\varphi}|^2\approx \underbrace{|A_+|^2-|A_-|^2}_{\text{calibrate to~} 0}+ 2\,\varepsilon\, \text{Re} (A_+ \bar B_+
     -A_- \bar B_-)
\end{equation}
By shifting the quad-detector up or down one can make the leading term vanish. This calibration makes the quad detector sensitive to the sub-leading term of $O(\varepsilon)$.

\section{Concluding remarks}

\begin{itemize}

     \item  The authors of \cite{LV} describe an experiment which agrees with a prediction of Aharonov-Vaidman theory \cite{AharonovVaidman} that post-selected quantum particles  have discontinuous quantum trajectories. 
     
     \item Orthodox quantum mechanicians will find the claim: Quantum particles have discontinuous paths, shocking. 
     
     \item In the Feynman path integral formulation of quantum evolution, \cite{Feynman}, a quantum particle tries all continuous paths simultaneously. Each path comes with a complex phase factor, $e^{iS/\hbar}$.

    \item The orthodox quantum mechanical analysis of the experiment \cite{LV} attributes the observation to the almost perfect destructive interference of paths from the source to the detector.

    \item Asher Peres coined the aphorism: Physics is not an exact science, it is the science of approximation. The observation in \cite{LV} reflects the approximate nature of the measurement whereas the statement: The photon path is discontinuous is dichotomic. As such, it is difficult to  reconcile with an approximation to reality. 
   
   \item As every amateur detective knows, the absence of footprints does not rule out a crime.
\end{itemize}
 \section*{Acknowledgment} I thank  Shimshon Bar-Ad and especially Lev Vaidman for several helpful conversations.  Yoav Sagi  for his criticism and  Oded Kenneth for insightful comments and pruning an early version of the manuscript from errors and obscurities. I thank Thierry Paul for careful reading of the manuscript and making many useful suggestions.

\printbibliography 

@article{LV,
  title = {Asking Photons Where They Have Been},
  author = {Danan, A. and Farfurnik, D. and Bar-Ad, S. and Vaidman, L.},
  journal = {Phys. Rev. Lett.},
  volume = {111},
  issue = {24},
  pages = {240402},
  numpages = {5},
  year = {2013},
  month = {12},
  publisher = {American Physical Society},
  doi = {10.1103/PhysRevLett.111.240402},
  url = {https://link.aps.org/doi/10.1103/PhysRevLett.111.240402}
}

@book{Feynman,
      author        = "Feynman, Richard Phillips and Hibbs, Albert Roach",
      title         = "{Quantum mechanics and path integrals}",
      publisher     = "McGraw-Hill",
      address       = "New York, NY",
      series        = "International series in pure and applied physics",
      year          = "1965",
      url           = "https://cds.cern.ch/record/100771",
}

@article{AharonovVaidman,
  title={The Two-State Vector Formalism: An Updated Review},
  author={Yakir Aharonov and Lev Vaidman},
  journal={Lecture Notes in Physics},
  year={2008},
  volume={734},
  pages={399-447}
}

@article{Controversy,
author={Vaidman, Lev},
title={Weak value controversy}, Journal={Philosophical Transactions of the Royal Society A: Mathematical, Physical and Engineering Sciences},
volume={375.2106},
year={2017},
}

@article{AP,
author = {Fuchs, Christopher and Peres, Asher},
year = {2000},
month = {09},
pages = {},
title = {Quantum Theory Needs No Interpretation"},
volume = {53},
journal = {Physics Today},
doi = {10.1063/1.883004}
}

\end{document}